\documentclass{aa}
\usepackage{graphicx}
\usepackage{txfonts}
\usepackage{color}

\begin{document}

\title{Magnetic flux transport on active cool stars and \\ starspot lifetimes}
\author{ E. I\c{s}\i k \and M. Sch\"ussler \and S. K. Solanki }

\institute{ Max-Planck-Institut f\"ur Sonnensystemforschung, Max-Planck-Str.
2,
37191 Katlenburg-Lindau, Germany \\
\email{[ishik,schuessler,solanki]@mps.mpg.de} }

\date{Received / Accepted }
\abstract
% Context
{Many rapidly rotating cool stars show signatures of large magnetic regions
at all latitudes. Mid-latitude starspots and magnetic regions have
characteristic lifetimes of 1 month or less, as indicated by observations
using (Zeeman-) Doppler imaging techniques. }
% Aims
{We aim to estimate the lifetimes of bipolar magnetic regions and starspots
on the surfaces of cool stars. We consider different possible configurations
for starspots and compare their flux variations and lifetimes based on a
magnetic flux transport model. }
% Methods
{We carry out numerical simulations of the surface evolution of bipolar
magnetic regions (BMRs) and magnetic spots on stars, which have radii and
surface rotational shears of \object{AB Doradus}, the Sun, and the \object{HR 1099} 
primary.
The surface flux transport model is based on the magnetic induction equation
for radial fields under the effects of surface differential rotation,
meridional flow, and turbulent diffusion due to convective flow patterns. We
calculate the flux evolution and the lifetimes of BMRs and unipolar
starspots, varying the emergence latitude, surface shear rate, and tilt
angle.}
% Results
{For BMRs comparable to the largest observed on the Sun, we find that varying
the surface flows and the tilt angle modifies the lifetimes over a range of one
month. For very large BMRs (area $\sim$10\% of the stellar surface) the
assumption of a tilt angle increasing with latitude leads to a significant
increase of lifetime, as compared to the case without tilt. Such regions can
evolve to polar spots that live more than a year. Adopting the observed
weak latitudinal shear and the radius of the active subgiant component of HR
1099, we find longer BMR lifetimes as compared to the more strongly sheared
\object{AB Dor} case. Random emergence of six additional tilted bipoles in an activity 
belt at $60^\circ$ latitude enhanced the lifetimes of polar caps up to 7 years. 
We have also compared the evolution and lifetime of monolithic
starspots with those of conglomerates of smaller spots of similar total area.
We find similar decay patterns and lifetimes for both configurations. }
% Conclusions
{} \keywords{Stars: magnetic fields -- magnetohydrodynamics (MHD) -- stars:
activity}

\titlerunning{Magnetic flux transport on active stars}
\authorrunning{E. I\c{s}\i k et al.}
\maketitle

%%%%%%%%%%%%%%%%%%%%%%%%%%%%%%%%%%%%%%%
\section{Introduction}
%%%%%%%%%%%%%%%%%%%%%%%%%%%%%%%%%%%%%%%

Magnetic flux emerges on the solar surface in the form of bipolar magnetic
regions (hereafter BMRs), the larger ones of which include sunspots, pores,
and plages (consisting of small-scale magnetic flux concentrations) in two
patches of opposite polarity. The observations indicate that, after their
emergence, magnetic flux in BMRs is subject to flux transport by means of
convective motions on different scales (granulation, supergranulation),
differential rotation, and meridional flow. Under the effects of these flows,
magnetic flux is redistributed on the surface, leading to flux cancellation
and polarity reversals of the polar field, which is possibly an important
ingredient of the underlying dynamo mechanism. Magnetic flux transport on the
surface of the Sun has been studied through numerical simulations by several
authors, who were able to reproduce the evolution of the radial magnetic
field on the solar surface through the magnetic cycle (e.g., Wang et al.
\cite{wns89a}, Dikpati \& Choudhuri \cite{dc95}, van Ballegooijen et al.
\cite{balle98}, Wang \cite{w98}, Mackay et al. \cite{mac02}; Baumann et al.
\cite{bsss04}).

Observations of spots and magnetic fields on stars other than the Sun,
observations are mainly restricted to rapidly rotating, i.e. active,
nearby G-K stars, because of the relatively high spectral resolution and low
noise required by the indirect surface imaging techniques. One of the striking observational results is that,
in contrast to the solar case, long-lived spots or spot groups
lie at high latitudes and often even cover
the rotational poles (see Strassmeier \cite{str02}), in some cases with
intermingling of opposite polarities (Donati et al. \cite{don03a}).
Such accumulation of magnetic flux at high latitudes can arise by a
combination of poleward deflection of rising flux tubes by the Coriolis force
(Sch\"ussler \& Solanki \cite{ss92}; Sch\"ussler et al. \cite{msch96};
Granzer et al. \cite{granzer00}) and surface flux transport after emergence
(Schrijver~\cite{sc01}; Schrijver \& Title~\cite{st01}).
Recently, Mackay et al. (\cite{mac04}) used a flux transport model to show
that the observed intermingling of large amounts of positive and negative
magnetic flux at very high latitudes could occur with a flux emergence rate
30 times that of the Sun, a range of emergence latitudes between
50-70$^\circ$, and a meridional flow of 100 $\mathrm{m~s^{-1}}$, which
is about 10 times faster than in the case of the Sun. Intriguingly,
the observed
fields of very active stars often appear to have a strong azimuthal component
(Donati et al. \cite{don03a}), which is not observed in the case of the 
Sun\footnote{Recently, a weak toroidal field component on the solar surface
has been reported, which varies in phase with the magnetic cycle (Ulrich \&
Boyden \cite{ulboy05}).}.

Observational techniques like Doppler imaging can only detect starspots which
are much larger than typical sunspots. Because of the limited resolution of
the imaging techniques, it is unknown whether the observed starspots are
single large spots or conglomerates of smaller spots (cf. Solanki \& Unruh
\cite{solun04}). In any case, except for the polar spots, which in some
cases persist up to decades (cf.~Hussain~\cite{hus02}), the observed
starspots
have lifetimes that are less than one month (Barnes et al. \cite{barnes98},
Hussain \cite{hus02}, K\H{o}v\'{a}ri et al. \cite{kv04}). An individual
sunspot (and presumably a starspot as well) is a more coherent structure than
the magnetic region in which it is embedded: because of the magnetic forces,
its magnetic flux is not passively transported by convective flows or sheared
by differential rotation, but resists to them (e.g., by suppressing
granulation). Sunspots gradually lose flux through (turbulent) erosion by
convection at the boundaries (Petrovay \& Moreno-Insertis \cite{pmi97}). The
decay time of an individual sunspot is proportional to its maximum diameter
(Petrovay \& van Driel-Gesztelyi~\cite{pv97}), which is consistent with the
turbulent erosion models (Petrovay \& Moreno-Insertis~\cite{pmi97}). On the
other hand, a cluster of small spots is more likely to be dispersed by
differential rotation, meridional flow, and supergranulation. The structure
and evolution of starspots, in comparison with sunspots, has been reviewed by
Schrijver~(\cite{s02}). So far there has been no theoretical investigation
devoted to the lifetimes of starspots.

In this study, we simulate BMR and starspot evolution on the basis of a
linear surface flux transport model, with an aim to explain the observed
starspot lifetimes, and to compare various configurations. We carry out
numerical simulations in order to examine the evolution of the radial
magnetic flux in monolithic as well as clustered forms, and to
infer lifetimes for starspots covering a range of sizes, initial
latitudes, and transport parameters. We apply the surface flux transport code
of Baumann et al. (\cite{bsss04}, see also Baumann \cite{bau05}) to the cases
of a single bipolar magnetic region with different parameters, to multiple 
BMRs at high latitudes, to a unipolar
cluster of starspots, and to a large monolithic, unipolar
spot. The plan of the paper is as follows: the surface flux transport model
is introduced in Sect.~\ref{sec:model} and the evolution of BMRs is
treated in Sect.~\ref{sec:bmrs}, with emphasis on the effects of surface
flows and the emergence latitude upon lifetimes. The results for
different starspot configurations are discussed in Sect.~\ref{sec:starspots}
and concluding remarks are given in Sect.~\ref{sec:conclusions}.

%%%%%%%%%%%%%%%%%%%%%%%%%%%%%%%%%%%%%%%
\section{Model setup}
\label{sec:model}
%%%%%%%%%%%%%%%%%%%%%%%%%%%%%%%%%%%%%%%

To simplify the problem, we ignore any horizontal components of the surface
magnetic field and assume that the magnetic field on the stellar surface is
directed only in the radial (vertical) direction, as well justified for the
case of
the Sun (Solanki \cite{sol93}, Martinez Pillet et al. \cite{mp97}). Stellar
magnetic regions are assumed to have a bipolar structure with a geometry
similar to those of the solar BMRs (except for
Sect.~\ref{sec:starspots}, in which unipolar spots are considered). The flux
transport model is restricted to the surface, which is defined by the stellar
radius, $r=R_{\star}$. The signed flux density of a BMR is written in the
form
\begin{eqnarray}
      \label{eq:net}
      B(\lambda,\phi) = B^+(\lambda,\phi) - B^-(\lambda,\phi),
\end{eqnarray}
where $\lambda$ denotes the stellar latitude and $\phi$ the longitude.
Following van Ballegooijen et al.~(\cite{balle98}) and Baumann et
al.~(\cite{bsss04}), we assume the unsigned field strength of the two
polarities of a newly emerged BMR to be
\begin{eqnarray}
      \label{eq:gaudist}
      B^{\pm}(\lambda,\phi) = B_0 \exp\left[- \frac{ 2\left[ 1-\cos\beta_{\pm}(\lambda,\phi) \right]}{\beta_{0}^2}\right].
\end{eqnarray}
Here, $ \beta_{\pm}(\lambda,\phi) $ are the heliocentric angles between any
given position $(\lambda,\phi)$ and the centre of the positive and negative
polarities, $(\lambda_{\pm},\phi_{\pm})$. The total size of a BMR is
controlled by $\beta_0$, the initial characteristic width of each polarity,
and by the angular separation of the centres of the two poles, $\Delta\beta$,
which are related by the relation $\beta_0 = 0.4\Delta\beta$. $B_0$ is
arbitrarily set to 250 G, a reasonable value for the case of the Sun
\footnote{ The magnetic field strength can be scaled arbitrarily, because the
field evolution described by Eq.~(\ref{eq:transport}) is linear and passive.
Therefore, a different peak strength for the initial field results in the
\emph{same} patterns of flux evolution, only scaled by a factor in the field
strength. Because we take a fixed fraction of $B_0$ as the threshold field
strength to define the size of an evolving BMR (see below), the resulting
flux evolution is independent of the initial peak strength. 
Another reason for choosing $B_0$=250~G is that we get magnetic fluxes 
comparable to those of solar active regions.}. 
Initially the BMR is placed at a given latitude according to one of the
following configurations: (1) both poles of the BMR are at the same latitude,
(2) the line joining the centres of the two polarities makes an angle
$\alpha$ with a latitudinal circle, which is called the tilt angle. The
latitude dependence is chosen to be $\alpha=0.5\lambda_0$, where $\lambda_0$
is the latitude of emergence. The orientation of a bipole axis (i.e.,
the line joining the centres of the two polarities) is such that the leading
polarity (with respect to the direction of rotation) is nearer to the
equator. The time evolution of the radial magnetic field is determined by the
induction equation, which is written in spherical coordinates
$(R_{\star},\lambda,\phi)$ as
\begin{eqnarray}
      \label{eq:transport}
      \frac{\partial B}{\partial t} & = & - \omega(\lambda) \frac{\partial B}{\partial \phi}
      + \frac{1}{R_{\star}\cos\lambda}\frac{\partial}{\partial\lambda} \left( \varv(\lambda) B\cos\lambda \right)\nonumber\\
      \noalign{\vskip 2mm}
      & & + \frac{\eta}{R_{\star}^2} \left[\frac{1}{\cos\lambda}\frac{\partial}{\partial\lambda} \left(\cos\lambda
      \frac{\partial B}{\partial\lambda} \right) +
      \frac{1}{\cos^2\lambda} \frac{\partial^2B}{\partial\lambda^2} \right]\nonumber\\
      \noalign{\vskip 2mm}
      & & + S(\lambda,\phi) - D_r(\eta_r),
\end{eqnarray}
where $\lambda$ is the latitude, $\omega(\lambda)$ is the angular
velocity of rotation, $\varv(\lambda)$ is the meridional flow velocity,
$\eta$ is the constant coefficient for horizontal turbulent diffusion
associated with the nonstationary convective motions (supergranulation in the
case of the Sun), $S(\lambda,\phi)$ is the source term describing the
emergence of the BMRs, and $D_r$ an additional decay term representing the
radial diffusion of magnetic flux with an effective diffusivity $\eta_r$
(Baumann et al. \cite{bss06}). For the numerical solution of
Eq.~(\ref{eq:transport}), the magnetic field is expanded in spherical
harmonics with a maximum degree of 63, which (for $R_{\star}=R_{\sun}$)
corresponds roughly to the observed size of supergranules ($\approx$30 Mm) on
the Sun. For the newly emerged BMRs, the spherical harmonic coefficients
defining the field distribution are multiplied by a spatial filter of the
form $\mathrm{exp}\big[-\beta_0^2 l(l+1)/4\big]$ (van Ballegooijen et al.
\cite{balle98}), in order to diminish the effect of ringing (Gibb's
phenomenon) caused by the truncation of the expansion in spherical harmonics.
\begin{figure}
% Fig.1
% transport.pro
      \centering
      \resizebox{\hsize}{!}{\includegraphics{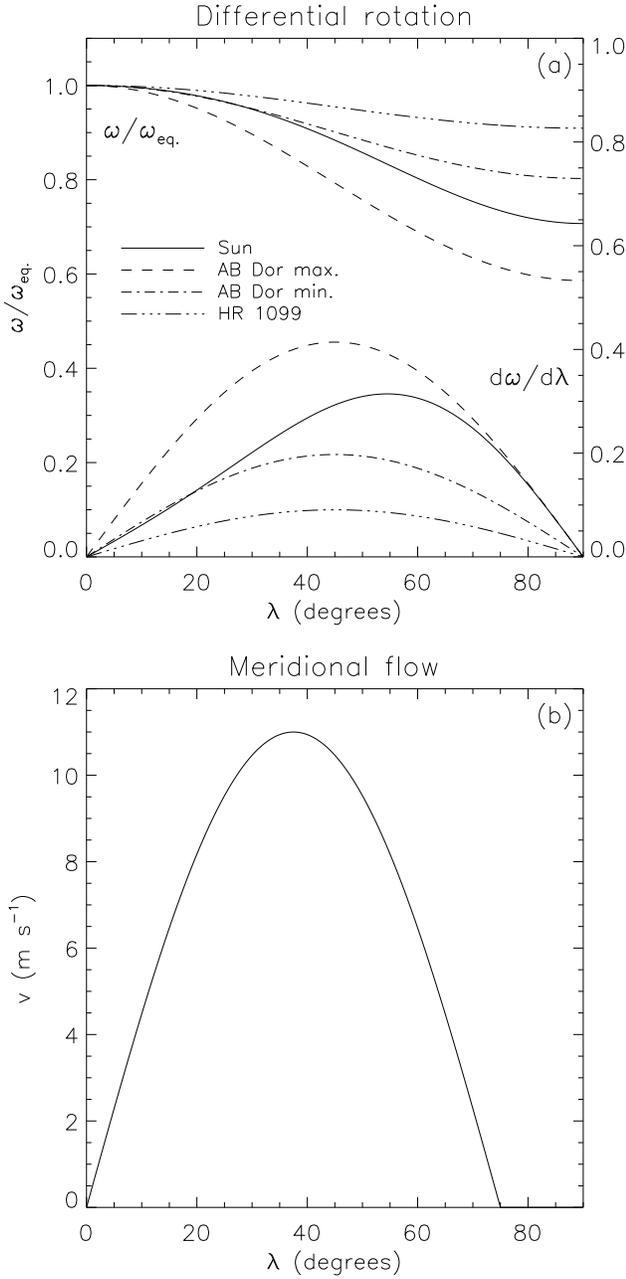}}
      \caption{(a) Angular velocity normalised to its equatorial ($\lambda=0$) value
      (left axis, top curves) and shear rate (right axis, bottom
      curves) for three stars: the Sun, \object{AB Dor}, and \object{HR 1099}.
      `AB Dor max.' and `AB Dor min.' correspond to the maximum and minimum observed
      $\Delta\omega$ values obtained by Donati et al. (\cite{don03b}).
      (b) Meridional flow velocity (positive means poleward flow). Both
      plots are for the northern hemisphere. The shear rate and the meridional flow
      profiles are antisymmetric with respect to the equator. }
      \label{fig:trans}
\end{figure}
\begin{figure}
% Fig.2
% polarx.pro
      \centering
      \resizebox{\hsize}{!}{\includegraphics{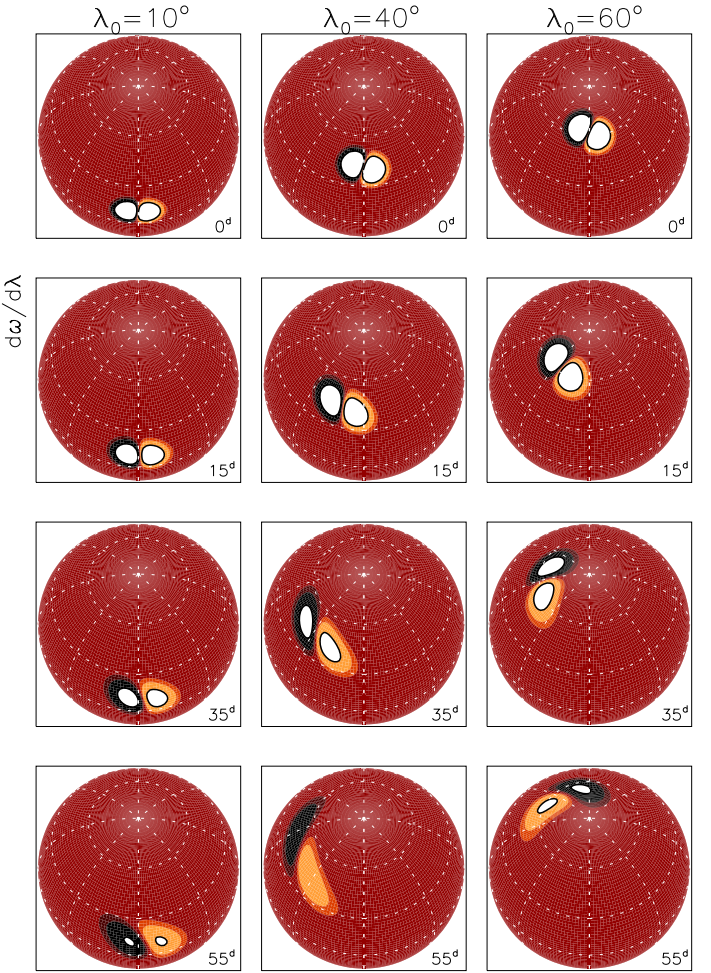}}
      \caption{Evolution of BMRs with different emergence latitudes ($\lambda_0$), with
      tilt angles equal to $0.5\lambda_0$, and with solar surface shear. 
      The two polarities indicated by black and yellow colour, and the regions with
      unsigned field strength above the threshold are represented by white regions
      surrounded by black contours. Contours for 4\% and 2\% of the initial peak
      strength are filled with shades of black and yellow in order to illustrate the
      overall shape of the region. 
      The projections are centred at a fixed meridian and $60^{\circ}$
      latitude. The latitude circles are plotted with 30$^\circ$ intervals. Elapsed
      time after emergence (in days) is indicated. Stellar (differential) rotation
      is shown with respect to the rest frame of the equator.
      The initial BMR configuration corresponds to
      $\beta_{0}=4^{\circ}, \Delta\beta=10^{\circ}$, leading to an initial area of
      about 323 square degrees, or 0.8\% of the solar/stellar surface area.}
      \label{fig:evo}
\end{figure}
\begin{figure}
% Fig.3
% evo1t.pro +life.pro
      \centering
      \resizebox{\hsize}{!}{\includegraphics{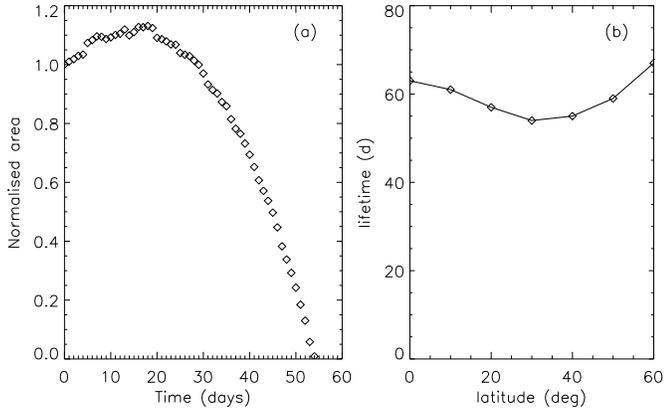}}
      \caption{(a) Time variation of BMR area normalised to its initial value, for a BMR
      with $\beta_{0}=4^{\circ}, \Delta\beta=10^{\circ}$ emerged at $40^\circ$.
      BMR area is defined as the area above a given threshold (see text). (b) BMR lifetime as a function of the emergence latitude.}
      \label{fig:evodr}
\end{figure}
\begin{figure}
% Fig.4
% life.pro
      \centering
      \resizebox{\hsize}{!}{\includegraphics{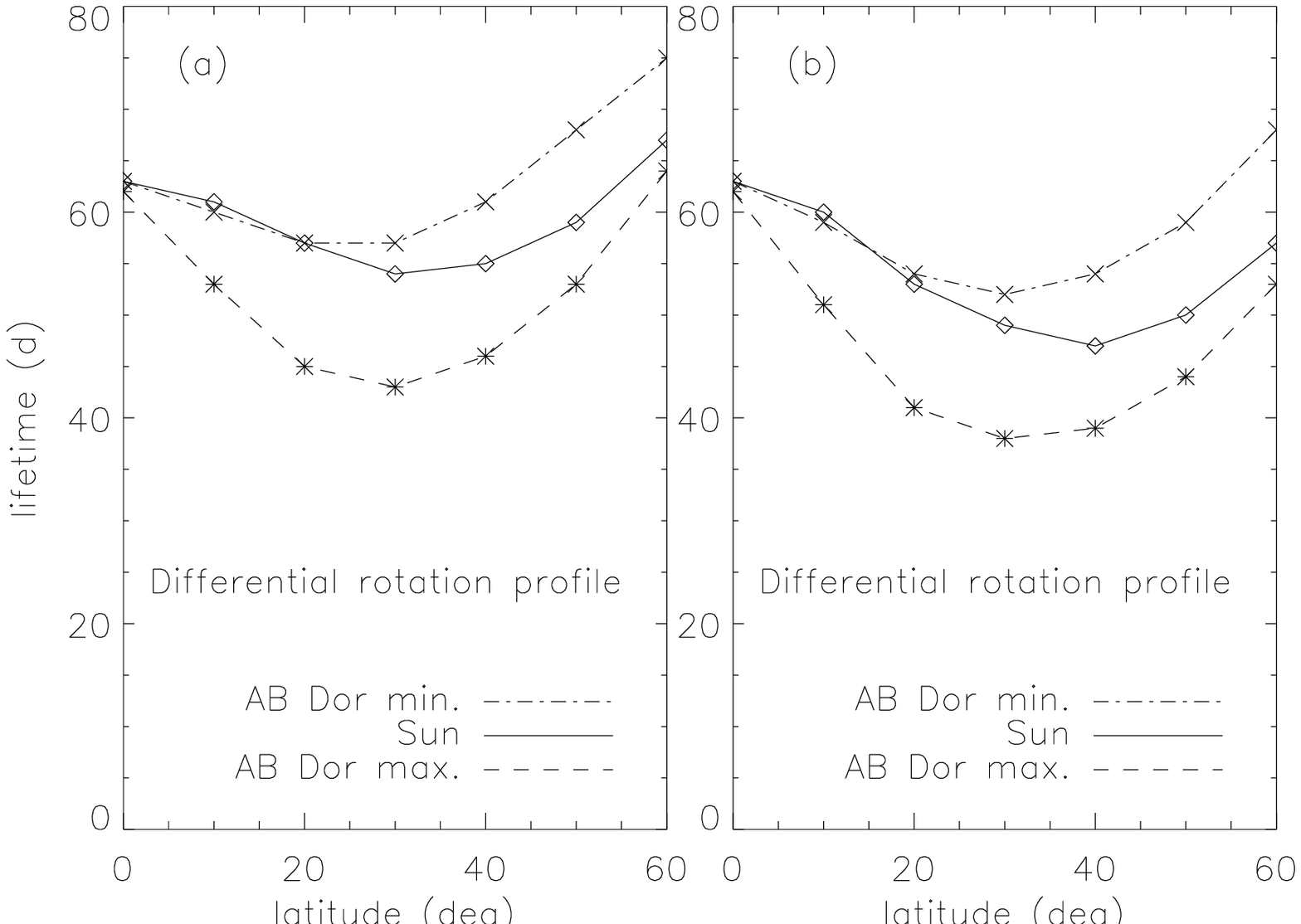}}
      \caption{Lifetime of BMRs as a function of the
      emergence latitude, for three differential rotation (surface shear)
      profiles: solar shear (diamonds), the minimum observed shear
      for \object{AB Dor} (crosses), and
      the maximum observed shear for \object{AB Dor} (asterisks). (a) for the
      tilt angle relation $\alpha=0.5\lambda_0$. (b) $\alpha=0$.}
      \label{fig:life}
\end{figure}

The relevant quantities for the flux transport are the latitudinal angular
velocity profile $\omega(\lambda)$, the meridional flow profile
$\varv(\lambda)$, the horizontal turbulent magnetic diffusivity, $\eta$, and
the effective radial diffusivity, $\eta_r$. As a reference configuration
we consider the values corresponding to the solar case. The
angular velocity profile is taken after Snodgrass (\cite{snod83}),
\begin{eqnarray}
\omega(\lambda) = 13.38 - 2.30\sin^2\lambda - 1.62\sin^4\lambda \;\;\;\;
\mathrm{deg~day^{-1}}, \label{eq:dr}
\end{eqnarray}
and the meridional flow profile (Snodgrass \& Dailey \cite{snodd96}; Hathaway
\cite{hath96}) is assumed as
\begin{eqnarray}
\varv(\lambda) = \left\{ \begin{array}{r@{\quad:\quad}l}-\varv_0\sin(\pi\lambda / \lambda_\mathrm{c}) & {\rm if}~|\lambda| < \lambda_\mathrm{c} \\
                              0 & {\rm otherwise},
                              \end{array} \right.
\label{eq:mf}
\end{eqnarray}
where $\varv_0=11~\mathrm{m~s^{-1}}$ and $\lambda_\mathrm{c}=\pm~75^{\circ}$
(van Ballegooijen et al.~\cite{balle98}; Baumann et al.~\cite{bsss04}). For
the case of rapid rotators, the following differential rotation profile is
used:
\begin{eqnarray}
\omega(\lambda) = \omega_{\mathrm{eq}} - \Delta\omega\sin^2\lambda.
\label{eq:dr1}
\end{eqnarray}
Here, $\omega_{\mathrm{eq}}$ is the equatorial angular velocity and
$\Delta\omega$ is the velocity difference between the poles and
the equator. The values of the latter are 5.54 and 2.64 deg
$\mathrm{day^{-1}}$, for \object{AB Dor} (maximum and minimum observed surface
shear rates), and 1.21 deg $\mathrm{day^{-1}}$ for \object{HR 1099} (Donati et al.
\cite{don03b}). The differential rotation and meridional flow profiles given
by Eqs.~(\ref{eq:dr})-(\ref{eq:dr1}) are shown in Fig.~\ref{fig:trans}.

In the following, we consider the evolution of BMRs on stars with various
shear rates and radii in order to investigate the impact of surface flows on
the lifetimes for BMRs of different sizes. Then we consider the evolution of
starspots in Sect.~\ref{sec:starspots}.

%%%%%%%%%%%%%%%%%%%%%%%%%%%%%%%%%%%%%%%
\section{Evolution of BMRs on stars}
\label{sec:bmrs}
%%%%%%%%%%%%%%%%%%%%%%%%%%%%%%%%%%%%%%%

In this section we present numerical simulations of single bipolar magnetic
regions and of multiple BMRs emerging randomly around a polar spot. Partly, BMRs are important because they harbour starspots
visible in Doppler imaging. Partly, the Zeeman-Doppler imaging technique
should be able to detect the largest BMRs on stars. In this connection, 
we consider starspots as bipolar magnetic regions. 
We take the horizontal diffusivity $\eta=600~\mathrm{km^2~s^{-1}}$
(Wang et al.~\cite{wns89b}) and the effective radial diffusivity
$\eta_r=100~\mathrm{km^2~s^{-1}}$ (Baumann et al.~\cite{bss06}). 
These values are appropriate for the Sun. 
We adopt them for all the simulations in this section, since the 
properties of the large-scale convective flows (which determine the 
turbulent diffusivities) are unknown for stars with a different mass 
in a different evolutionary stage. This uncertainty should be kept in mind 
when considering the results for stars other than the Sun. 
The smallest
BMR considered here has $\beta_{0}=4^{\circ}, \Delta\beta=10^{\circ}$,
roughly representing the largest solar BMRs shortly after their complete
emergence, and also the smallest structures that can be resolved in
Zeeman-Doppler reconstructions of active cool stars (Donati et al.
\cite{don03a}). We consider this angular size as the `standard' one in
Sec.~\ref{sec:abdor} and Sec.~\ref{sec:hr1099}. The area of a BMR at a given
time is determined from the corresponding simulated field distribution on an
angular grid with 360 points in azimuth and 180 points in latitude, by adding
up the area for which a threshold magnetic field strength is exceeded. The
latter is taken to be 0.14 times the value of the initial peak field strength
$B_0$, so that the initial area for the 'standard' BMR is equal to 323 square
degrees, or about 8 thousands of the stellar surface area. The threshold field
strength was determined by requiring that the lifetime of a BMR (as
defined by the time when the field strength falls entirely below the
threshold) with $\beta_{0}=4^{\circ}, \Delta\beta=10^{\circ}$ becomes equal
to two months, which is about the lifetime of a similar-sized BMR
around a solar minimum, which we estimated visually from SOHO/MDI
synoptic magnetograms for the time period 25.06.2004 and 14.09.2004 
(Carrington rotations from 2018 to 2020). 
We obtained this estimate by tracking relatively
isolated BMRs of size $\Delta\beta\approx 10^{\circ}$ at
a latitude of about $10^\circ$ on consecutive synoptic
magnetograms. The estimated
lifetimes of around two Carrington rotations, roughly two
months after their maximum size of development, can be compared with the
largest BMRs shown by Harvey~(\cite{h}), which have areas up to 70 square
degrees and lifetimes up to 3 months. 

%%%%%%%%%%%%%%%%%%%%%%%%%%%%%%%%%%%%%%%%%%%%%%%%
\subsection{Main sequence star of solar radius}
\label{sec:abdor}
%%%%%%%%%%%%%%%%%%%%%%%%%%%%%%%%%%%%%%%%%%%%%%%%

In this section we present simulations of the evolution of single BMRs on a
star of solar radius. Examples are the Sun and the rapidly rotating K0-dwarf
\object{AB Doradus}, ($R\approx R_{\sun}$, according to Ambruster et
al.~\cite{amb03}). Fig.~\ref{fig:evo} illustrates the dependence 
of BMR lifetime on emergence latitude. It shows the evolution of BMRs with
$\beta_{0}=4^{\circ}, \Delta\beta=10^{\circ}$, for three different emergence
latitudes: $10^\circ$, $40^\circ$, and $60^\circ$. The flow profiles
represent the solar case given by Eqs.~(\ref{eq:dr}) and (\ref{eq:mf}).
It can be seen that, 55 days after the emergence, the BMR emerged at 
$\lambda_0=60^\circ$ has more remaining flux than the one at $\lambda_0=10^\circ$, 
while the one at $\lambda_0=40^\circ$ has already lost all its flux 
with field strengths above the threshold. 
Fig.~\ref{fig:evodr}a shows the evolution of
area (normalised to its initial value) for the BMR at 
$\lambda_0=40^\circ$ 
and Fig.~\ref{fig:evodr}b shows the BMR lifetime as a function of the emergence
latitude. For all emergence latitudes, the area increases in the initial
phase by about 10 percent, owing to the spreading by diffusion of the field
above the threshold (e.g., Fig.~\ref{fig:evodr}a for $40^\circ$). After
about 15 days, the region above the threshold field strength begins to shrink, as
the reduction of its length scales (steepening of the field gradient) by means of 
surface shear leads to increasing diffusion rates. The strength of this
effect is proportional to the local surface shear (see
Fig.~\ref{fig:trans}a), and is partly responsible for the variation of lifetime as a
function of emergence latitude (Fig.~\ref{fig:evodr}b). The effect can also
be seen when one compares the evolution of BMRs at $10^\circ$ and $40^\circ$
in Fig.~\ref{fig:evo}. On the other hand, the bipole at $60^\circ$ lives
longer than the ones at $10^\circ$ and $40^\circ$, although the local shear
rate is higher for the former. The reason is the meridional flow (see
Fig.~\ref{fig:trans}b): firstly, there is an acceleration of the poleward
flow with increasing latitude until around $37^\circ$, above which the flow
is decelerated. Therefore, a BMR emerging at a latitude lower (higher) than
about $37^\circ$ experiences a diverging (converging) flow. Secondly,
low-latitude BMRs are advected to latitudes with stronger shear, whereas the
high-latitude BMRs are moved to regions with less shear. Both of the effects
mentioned above contribute to the longer lifetimes of high-latitude BMRs.

Next we consider the effect of varying the latitudinal rotation profile
(surface shear) by using the results for the solar profile (as above),
and those for \object{AB Doradus} (Fig.~\ref{fig:trans}a), with observed minimum
and maximum values for $\Delta\omega$ of 2.64 and 5.54 deg
$\mathrm{day^{-1}}$, respectively (Donati et al.~\cite{don03b}). The
variation of BMR lifetime as a function of the emergence latitude for the
three cases is shown in Fig.~\ref{fig:life}, for tilt angles
$\alpha=0.5\lambda_0$ (Fig.~\ref{fig:life}a) and $\alpha=0$
(Fig.~\ref{fig:life}b), respectively. The lifetimes are affected by the shape
of the rotational shear profile (cf. Fig.~\ref{fig:trans}a). For instance,
for $\lambda_0=30^\circ$, 56\% weaker shear leads to an about 37\% (14
days) longer lifetime. The effect of meridional flow becomes more noticeable
for the cases with smaller rotational shear. For $\alpha=0$, the BMRs suffer
more flux cancellation at the longer neutral line between the opposite
polarities, so that they live shorter than their tilted counterparts. The
presence of a tilt angle $\alpha=0.5\lambda_0$ prolongs the lifetimes,
particularly so at higher latitudes: because the opposite polarities have
different angular velocities, they gradually separate so that the neutral
line at the interface for flux cancellation shortens (see
Fig.~\ref{fig:evo}).

\begin{figure}
% Fig.5
      \centering
      \resizebox{\hsize}{!}{\includegraphics{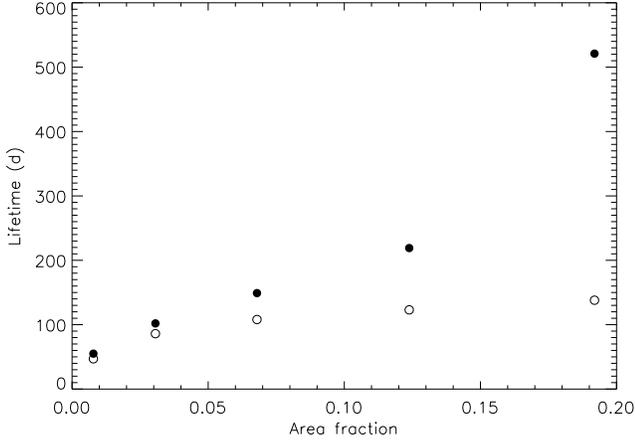}}
      \caption{Variation of BMR lifetime as a function of initial size. The initial angular
      radii of each pole of the BMR are $\beta_0=4^{\circ},8^{\circ},12^{\circ},
      16^{\circ},20^{\circ}$, and the initial separation between the poles satisfies the
      relation $\Delta\beta=2.5\beta_0$. Open circles denote the case with tilt
      angle $\alpha=0$, filled circles with $\alpha=0.5\lambda_0$.}
      \label{fig:lifebig}
\end{figure}

\begin{figure}
% Fig.6
      \centering
      \resizebox{\hsize}{!}{\includegraphics{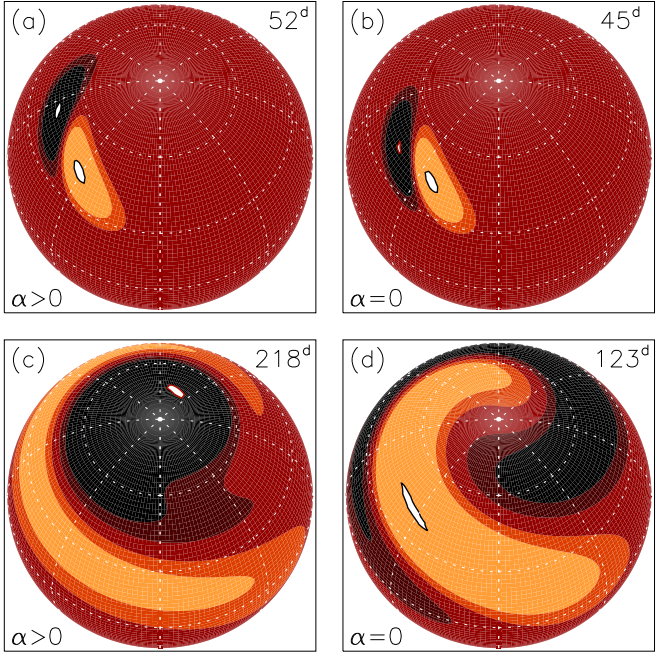}}
      \caption{Magnetic field distributions for relatively small (a,b) and
      very large (c,d) BMRs (area fractions 0.008 and 0.12 of the stellar surface area),
      emerged at $40^{\circ}$ with tilt (a,c) and without tilt (b,d). The snapshots
      for each of the four cases represent the moment shortly before the peak field
      strength falls below the threshold. The number of days since the bipole emerged is given in each frame. 
      Colours and contours are the same as in Fig.~\ref{fig:evo}. Non-zero tilt yields longer lifetimes, particularly so for a very large BMR
      which eventually forms a polar cap (panel c). }
      \label{fig:polar2}
\end{figure}

\subsection{Size dependence of lifetime}
\label{sec:size}

Larger BMRs have systematically longer lifetimes owing to a larger initial
area. Fig.~\ref{fig:lifebig} shows the lifetimes, for $40^{\circ}$ emergence
latitude, of BMRs with initial area equal to or larger than in the reference
case presented above. The presence of a tilt (filled circles)
leads to polarity separation by differential rotation, thus reducing the
cancellation rate and leading to longer lifetimes, particularly so for very
large BMRs. This is illustrated in Fig.~\ref{fig:polar2}, which shows BMRs of
initial area fractions (area covered by the BMR as a fraction of the stellar 
surface area) 0.008 and 0.12 shortly before completely falling below
the threshold. For non-zero tilt ($\alpha > 0$), the follower polarity of the large BMR
forms a polar cap after about 6 months (panel c). Its nearly circular shape
decelerates its diffusive spreading because the shear can no longer reduce
its effective length scale. For the non-tilted case ($\alpha > 0$), however, the two
polarities rotate with the same speed and spiral towards the rotational pole
through the action of the meridional flow, leading to enhanced flux
cancellation at the elongated neutral line.

\subsection{The case of \object{HR 1099}}
\label{sec:hr1099}

In a subgiant star like the active component of the binary system \object{HR 1099}
with a radius of $3.3$ $R_{\sun}$, lifetimes for the same initial area
fractions as in the case of a 1 $R_{\sun}$ star are much longer.
This is expected for two reasons: firstly, surface differential rotation is weaker in
the case of \object{HR 1099} (see Fig.~\ref{fig:trans}), and secondly, the
characteristic decay time $\tau_l$ of the eigensolution of the diffusion equation is
given by
\begin{equation}
      \tau_l = \frac{R_{\star}^2}{\eta l(l+1)},
      \label{eq:lifetime}
\end{equation}
so that for features of the size corresponding to the spherical harmonic degree 
$l$, $\tau_l$ 
scales with $R_{\star}^2$. The larger the radius of the star, the longer the
lifetime of regions with the same area fraction.
The presence of flows modifies the simple quadratic dependence of decay time
on the stellar radius in Eq.~(\ref{eq:lifetime}), and prevents us from simply
rescaling. We substitute the radius of the active component of \object{HR 1099} (3.3
$\mathrm{R}_{\sun}$), its observed differential rotation rate, ($\Delta\omega=1.21~\mathrm{deg~day^{-1}}$), and the solar-like
meridional flow (Eq.~\ref{eq:mf}) into the flux transport simulation.
The imposed BMR emerges at a latitude of $40^\circ$; its initial fractional area is 0.008 (323 square degrees), and the tilt angle is 
$20^\circ$. The resulting lifetime is about 290 days ($\approx$ 9.5 months), 
which must be compared
to the value of 55 days ($\approx$ 2 months) for $R=R_{\sun}$ for the same fractional area and the solar differential rotation profile. Scaling with ${R_{\star}^2}$ would give $55 \times 3.3^2 \approx 600$ days (1.6 years). The difference between the calculated 
lifetime and the diffusion time scale is due 
mainly to the surface differential rotation, which reduces the characteristic 
length scales of the BMR. Therefore, the rate of diffusion increases with 
time, compared to the constant rate in the absence of large-scale flows. 

\begin{figure}
% Fig. 7
      \centering
      \resizebox{\hsize}{!}{\includegraphics{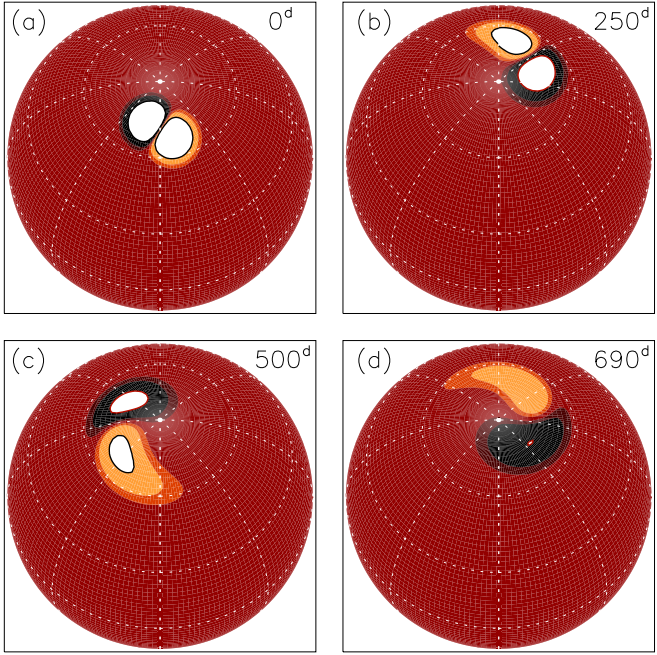}}
      \caption{Evolution of a polar BMR with  $\beta=4^\circ,\Delta\beta=10^\circ$,
      starting at $\lambda_0=70^\circ$ on a star with a radius of $3.3R_{\sun}$.
      The effects of weak differential rotation
      ($\Delta\omega=1.21~\mathrm{mrad~day^{-1}}$) and the same
      meridional flow profile and magnetic diffusivity as in the solar case
      are considered. The large stellar radius and the weak surface shear at high latitudes of the star lead to a lifetime of about two years. Since the meridional flow ceases at $75^\circ$, The BMR is not completely drifted up to the rotational pole.}
      \label{fig:polarg}
\end{figure}

For a BMR of the same size, but starting at $\lambda_0=70^\circ$, the
evolution with the same transport parameters leads to a lifetime of 
about 2 years, as shown in Fig.~\ref{fig:polarg}. Therefore, the long lifetimes
of polar spots on RS CVn-type active stars can possibly be related to the low
rate of shear at the polar regions of (sub)giant stars and a
poleward-decelerating meridional flow, particularly if the BMRs emerge 
at high latitudes and are strongly tilted.

\subsection{Emergence of multiple bipoles}
\label{sec:more}

We have shown in Sec.~\ref{sec:size} that when a polar cap of one polarity 
is developed, its decay is significantly slower than for the lower 
latitudes. The example case shown in Fig.~\ref{fig:polar2}c yields a 
lifetime of about 7 months. 
Now, in the course of its evolution, we let six more bipoles of fractional area 
0.03 to emerge at a latitude of $60^\circ$, at random longitudes and 
random times within a period of 500 days (1.4 years). 
This 
high-latitude temporary "activity belt" extends the lifetime of the polar spot to 
about 2.4 years (see the supplementary animation file {\tt dwarf.mpg}). 
The enhancement of lifetime by means of newly emerging bipoles does not depend 
on when the random emergence period begins within the lifetime of the 
already existing polar BMR. 

In the case of the subgiant star of Sec.~\ref{sec:hr1099}, the lifetime of 
a polar spot is a factor of about $(R_{\star}/R_{\sun})^2$ longer than the one 
having the same fractional area on a solar-radius star. The lifetime of a 
unipolar polar cap of fractional area 0.004 ($\beta_0=4^\circ$) is about 3 years 
on the subgiant star, compared to about 0.3 years on the dwarf. 
Random emergence of six bipoles of fractional area 0.03 during 
500 days (1.4 years) at $\lambda_0$=$60^\circ$ prolongs 
the lifetime of the polar spot to about 10 years (see the supplementary 
animation file {\tt subgiant.mpg}). 

%%%%%%%%%%%%%%%%%%%%%%%%%%%%%%%%%%%%%%%
\section{Evolution of starspots}
\label{sec:starspots}
%%%%%%%%%%%%%%%%%%%%%%%%%%%%%%%%%%%%%%%
It is not known whether the observed starspots are monolithic structures or
conglomerates of smaller spots. In addition, the only general information
available regarding their evolution are lifetime estimates, indicating values
of less than one month at mid-latitudes of rapid rotators
(Hussain~\cite{hus02}). Here we compare different possible configurations for
starspots or starspot groups of sizes comparable to those observed at
mid-latitudes of rapid rotators.
Sunspots and their clusters with relatively long lifetimes are unipolar
features. Therefore, in contrast to the BMR simulations presented above, we
now consider starspots to be unipolar regions, with the other polarity placed
on the opposite hemisphere, in order to conserve the total flux on the
surface. We also assume that the diffusion rate of a starspot is reduced
compared to the case of a BMR, which consists of spots, plages, and ephemeral
regions. The reason is that the strong and coherent magnetic fields in a
starspot can suppress convection, as in the case of sunspots. This effect is
represented in our simulations by choosing a magnetic diffusivity that is
much lower than the value adopted for BMR evolution. The observed decay rates
of sunspots correspond to a diffusivity of 10-50 $\mathrm{km^2~s^{-1}}$
(Martinez-Pillet et al.~\cite{marpi93}). We adopt a value of $\eta =
50~\mathrm{km^2~s^{-1}}$ for the starspot simulations. In the following, we
describe simulated scenarios in the presence and absence of large-scale flows
in order to discriminate between the effects; we consider monolithic and
cluster structures as well as two BMR-like models to compare their evolution.
All configurations harbour the same amount of total magnetic flux
($1.52\times10^{22}$ Mx). The cases are described in Table~\ref{table:1}.

\begin{table*}
\caption{Configurations for starspots}
\label{table:1}
\centering
\begin{minipage}{2\columnwidth}
\centering
\begin{tabular}{c c c c}
\hline\hline
Configuration & Fractional area & Large-scale flows\footnote{Solar-like 
differential rotation and meridional flow ($-$ if absent, $+$ if present).} 
& $\eta$ (km$^2$~s$^{-1}$) \\
\hline
Monolithic spot & 0.005 & $-$ & 50 \\
Monolithic spot \& flows & 0.005 & $+$ & 50 \\
Cluster of large spots & 0.006 & $-$ & 50 \\
Cluster of large spots \& flows & 0.006 & $+$ & 50 \\
Cluster of small spots\footnote{The spots are not individually followed, 
but represented by a weak (25 G) unipolar magnetic field under the 
influence of large-scale flows.} & 0.05\footnote{Including both magnetic 
and non-magnetic regions} & $+$ & 50 \\
Passive BMR\footnote{A completely passive flux distribution, similar to 
the BMR simulations shown in Sect.~\ref{sec:bmrs}.} & 0.05 & $+$ & 600 \\
\hline
\end{tabular}
\end{minipage}
\end{table*}

\begin{figure}
% Fig. 8
% evo1tspots.pro
      \centering
      \resizebox{\hsize}{!}{\includegraphics{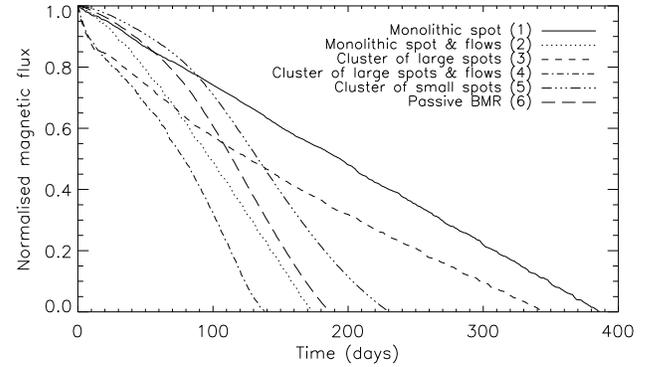}}
      \caption{Evolution of magnetic flux of the area above the threshold field
      strength of $0.14B_0$ normalised to the initial value ($\sim10^{22}$ Mx), for six
      different flux    configurations, all of which are started at $\lambda_0=50^{\circ}$
      (see main text). }
      \label{fig:conmon}
\end{figure}
\begin{figure}
% Fig. 9
      \centering
      \resizebox{\hsize}{!}{\includegraphics{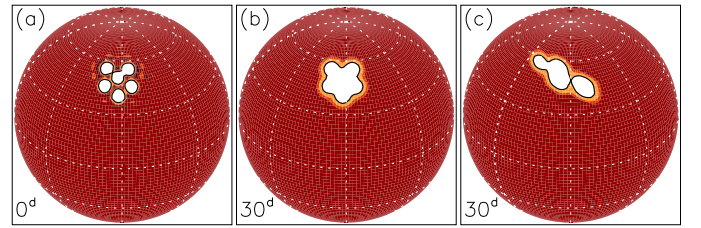}}
      \caption{Magnetic field evolution for a cluster of unipolar magnetic regions,
      with $\eta=50$ $\mathrm{km^2~s^{-1}}$ (cases 3 \& 4). (a) The initial field distribution,
      (b) the field distribution 30 days after the emergence, without large-scale flows (case 2),
      (c) the field distribution 30 days after the emergence, with solar-like large-scale flows
      (case 3).}
      \label{fig:polar3}
\end{figure}

The threshold field strength which determines the observable flux is taken to
be 0.14 times the initial peak field strength.
Fig.~\ref{fig:conmon} shows the evolution of magnetic flux at a field
strength above the threshold, normalised to its initial value. In case 1, the
flux decays nearly linearly in time, while for case 2 (the same monolithic
spot, but with large-scale flows) the decay is not linear in time,
because the surface shear and the meridional flow modify the length scale of
the spot. For large-spot cluster cases, with and without flows (cases 3 \&
4), the initial decay is much faster than in the other cases: the
conglomerate structure contains smaller flux elements, which diffuse faster
than the group as a whole. The situation is shown in
Fig.~\ref{fig:polar3}: For case 3, once the spots coalesce to a more
diffuse patch, the effective length scale becomes larger so that the decay
rate is reduced and the subsequent evolution takes place largely with the
same rate as for the circular, monolithic region (case 1). When flows are
introduced (case 4), the cluster of large spots decays much faster owing to
the decrease of length scale by differential rotation. For cases 5 \& 6,
the evolutions are similar to each other, since they differ only in 
magnetic diffusivity. In general, the effect of flows for the cases 2 and
4-6 leads to shorter lifetimes and nonlinear flux decay (for field strength
above the threshold), whereas the lack of flows leads to longer lifetimes and
linear decay for the cases 1 and 3.

The linear decay of flux for the monolithic spot without flows can be
understood through a simple analytical model. Consider the diffusion of a
scalar field $B$, which is initially distributed with axial symmetry on a
plane. For simplicity we neglect the curved surface on the sphere, which is
appropriate as long as $R_{\mathrm{spot}}\ll R_{\star}$. We write the
diffusion equation in cylindrical polar coordinates $(r, \theta, z)$ for an
axisymmetric field of strength $B$ along the $z$-direction as
\begin{eqnarray}
      \label{eq:diff}
      \frac{\partial B}{\partial t} & = &
      \eta~\left(\frac{\partial^2 B}{\partial r^2}+
      \frac{1}{r}\frac{\partial B}{\partial r}\right),
\end{eqnarray}
where $\eta$ is the magnetic diffusivity. Assuming an initial field with a
Gaussian profile,
\begin{eqnarray}
      \label{eq:init}
      B(r,t=0) = B_0~\mathrm{exp}\left(\frac{-r^2}{R_0^2}\right),
\end{eqnarray}
a self-similar solution of the following form can be written as
\begin{eqnarray}
      \label{eq:sol}
      B(r,t) = \frac{\Phi_{0{\rm T}}}{\pi R^2(t)}\mathrm{exp}\left(\frac{-r^2}{R^2(t)}\right).
\end{eqnarray}
Here $\Phi_{0\rm T} = \int_0^{\infty}\int_0^{2\pi}B(r,0)rd\theta dr = \pi
R_0^2 B_0$ is the total magnetic flux as the conserved quantity and $R(t) =
(4\eta t+R_0^2)^{1/2}$ represents the characteristic length scale, which
varies with time, so that the diffusion time-scale $\eta/R^2(t)$ also is a
function of time. Integrating the flux density of the area with field
strength above a threshold level $f$ (taken as 0.14 in our simulations), we
find the time-dependent flux above the threshold as
\begin{eqnarray}
      \label{eq:flux}
      \Phi(t) = \Phi_0~\left(1 - \frac{4f\eta t}{(1-f)~R_0^2}~\right),
\end{eqnarray}
where $\Phi_0=\Phi_{0{\rm T}}(1-f)$ is the initial value. The linear decrease
of the flux in Eq.~(\ref{eq:flux}) is reproduced in the numerical simulation
of the monolithic spot case 1 in Fig.~\ref{fig:conmon}. Taking the
initial characteristic radius $R_0$ as the width $\sigma$ of the
initial Gaussian profile of the numerical simulation,
the lifetime of the enclosed region is 1623 days (about 1.5 diffusion time
scales) from Eq.~\ref{eq:flux}, in good agreement with the numerical
simulation for $\eta_{\rm r}=0$.

%%%%%%%%%%%%%%%%%%%%%%%%%%%%%%%%%
\section{Discussion and conclusions}            %
\label{sec:conclusions}       %
%%%%%%%%%%%%%%%%%%%%%%%%%%%%%%%%%
We have made numerical simulations of the surface transport of bipolar
magnetic regions and starspots in various configurations, all regarded as
purely radial magnetic fields with Gaussian initial profiles. 
Lifetimes calculated here are for isolated bipolar regions and 
unipolar spots or spot groups. In general, emergence of other bipoles 
at or around the location of an existing bipolar region can lead to a 
complete change of the spot distribution, which can be interpreted as 
the disappearance of a spot. This means that the real lifetimes should be 
lower to some extent which is determined by the emergence rate of BMRs.
The estimated
lifetimes of mid-latitude BMRs are in accordance with the estimated decay
time of stellar active regions based on Ca~\ion{II} H \& K observations
(Donahue et al.~\cite{doh97}) and Doppler imaging of rapid rotators (Barnes
et al.~\cite{barnes98}, Hussain~\cite{hus02}, K\H{o}v\'{a}ri et
al.~\cite{kv04}). Very low correlations between observations recorded a month
apart set an upper limit for the lifetimes of low- to mid-latitude magnetic
regions on rapid rotators (see Hussain \cite{hus02}). When considered
together with our simulations (assuming that the transport parameters for a
rapidly rotating G-K dwarf are similar to the solar values), the observed
lifetimes of 1 month also set an upper limit on the \emph{sizes} of BMRs to
be $10^\circ$-$20^\circ$ in angular diameter, or about 1\% of the total
surface area for a star with $R=R_{\sun}$. Our simulations for such a star
indicate that the differential rotation within the observed range of values for
\object{AB Dor} does not have a very
significant effect on BMR lifetimes. Varying the surface shear or the
emergence latitude changes the lifetimes by less than a month. 
We note that the simulations which were run without the meridional flow 
has shown that the absence of such a flow shortens the lifetime of a 
high-latitude BMR for only a few days. Therefore, the 
meridional flow does not disturb the above mentioned result.
The differences in lifetimes for different emergence latitudes are caused by
the flow pattern on the surface:
\begin{enumerate}
\item Stronger surface shear in mid-latitudes lead to 
lifetimes that are about 20 days ($\approx~30\%$) shorter than those for 
low ($10^\circ$) and high ($60^\circ$) latitudes. 
\item The sign and the amplitude of the gradient of the meridional
flow velocity also has an effect upon lifetimes. The presence of a diverging
flow at low latitudes and a converging flow at high latitudes lead to
shorter and longer lifetimes, respectively.
In addition, meridional flow advects low-latitude BMRs to regions
with stronger shear, thus the lifetime is shortened. It also advects the
mid-latitude BMRs to higher latitudes with weaker shear, and this 
prolongs the lifetime. 
For instance, the lifetime of a BMR emerged at $60^\circ$ latitude 
is 15 days longer than the one emerged at $30^\circ$ latitude. This is 
the case for both of the \object{AB Dor} surface shear values. Hence, high-latitude BMRs live longer than low-latitude ones, even when the local surface shear 
is the same. In our model, this is caused only by the meridional flow.
\end{enumerate}

BMRs which are larger than those discussed above live longer, and this
tendency is substantially increased when a tilt angle according to the
relation $\alpha=0.5\lambda_0$ is assumed. For instance, a polar cap forms
about six months after the emergence at $40^\circ$ of a BMR initially
covering 12\% of the stellar surface. Once such a ``polar spot'' is formed,
it will not be significantly affected by differential rotation because of its
roughly circular shape and circumpolar boundaries. When there 
are more BMRs 
with the same polarity orientation, emerging at e.g. $60^\circ$ latitude, 
we have shown that the lifetime of the polar spot is prolonged significantly.

For (sub)giant stars we find that BMRs emerging at high latitudes can persist
as polar spots for more than 2 years. This is caused by the large radius
of the star and the meridional flow pattern, which is assumed to be
decelerating with latitude for $37^\circ < \lambda < 75^\circ$. Furthermore,
during the evolution of a polar cap, emergence of new BMRs at high latitudes
may inject fresh flux of follower polarity and this can lead to a
longer persistence of the polar cap. This can possibly explain
the observed polar spots with very long lifetimes (cf. Hussain~\cite{hus02}).

In summary, we suggest that the observed long lifetimes for polar spots in both 
dwarf and giant stars are likely to be caused by
\begin{enumerate}
\item high-latitude emergence of BMRs, as indicated by the numerical 
simulations of rising flux tubes (Sch\"ussler et al.~\cite{msch96}, Granzer et al.~\cite{granzer00}), 
\item supply of follower-polarity flux by transport from mid-latitudes,
\item weak differential rotation near the poles, and its inefficiency
in disrupting polar magnetic regions,
\item the possibility of small turbulent magnetic diffusivity, 
owing to larger supergranules in subgiant atmospheres. 
\end{enumerate}
The possibility (2) has been considered by Schrijver \& Title~(\cite{st01}),
whose simulations show the formation of a unipolar ring of spots surrounding
an opposite-polarity polar cap on a highly active solar analogue. Process (1)
was also studied in the simulations by Mackay et al.~(\cite{mac04}),
resulting in the intermingling of opposite polarity regions near the poles,
owing to emergence at high-latitudes and an assumed meridional flow amplitude
about 10 times larger than the solar value.

It is not obvious how to describe starspots and spot clusters in the
framework of a linear surface flux transport model. Extrapolations of spot
areas to stars more active than the Sun (Solanki \& Unruh~\cite{solun04})
indicate that a large fraction of the observed starspots are smaller than the
resolution limits of Doppler imaging maps and thus might be missed on
existing reconstructions. These authors further suggest that, provided the
spot areas are lognormally distributed, it is likely that the observed spots
on RS CVn stars are actually conglomerates of smaller spots. Our simulations
of starspots indicate that a large monolithic spot (case 1 in
Sec.~\ref{sec:starspots}) and a similar-sized cluster of large spots (case 3)
have similar lifetimes. Thus they do not favour either of the two configurations. 
In the cases 1 and 3, the magnetic flux of the
region above the threshold field strength decreases linearly with time. We
have demonstrated that this is a natural consequence of the two-dimensional
diffusion of a scalar quantity with a Gaussian initial profile, when a region
above the detection threshold of the quantity is considered.

We have treated the configuration and transport of a unipolar spot/cluster in
a number of other ways, all chosen such that the total flux is kept the same.
As a variant of case 3, we considered the effects of solar-like differential
rotation and meridional flow on a cluster of large spots (case 4). The
resulting lifetime turned out to be strongly reduced in comparison to the
case without large-scale flows. In the initial phase, the individual spots
expand rapidly because of their relatively small sizes. After their
coalescence, the general evolution pattern becomes similar to that of the
case 2, in which the evolution of a monolithic spot is studied in the presence of 
large-scale flows.
Furthermore, we have considered a region (case 5) with an area about 10
times larger than a monolithic spot having the same magnetic flux. 
This is a very simple representation of a cluster of
small starspots on a large area was implicitly assumed. This configuration
has an intermediate lifetime, between that of the large-spot cluster with
flows (case 4) and of the one without flows (case 3). In summary, the
numerical simulations presented in Sec.~\ref{sec:starspots} indicate that the
lifetime of a unipolar spot or spot cluster can differ by a factor of about 2.5
depending on the assumed initial flux configuration and the effects of
large-scale flows. 

\begin{figure}
      \centering
      \resizebox{\hsize}{!}{\includegraphics[bb=30 14 494 337]{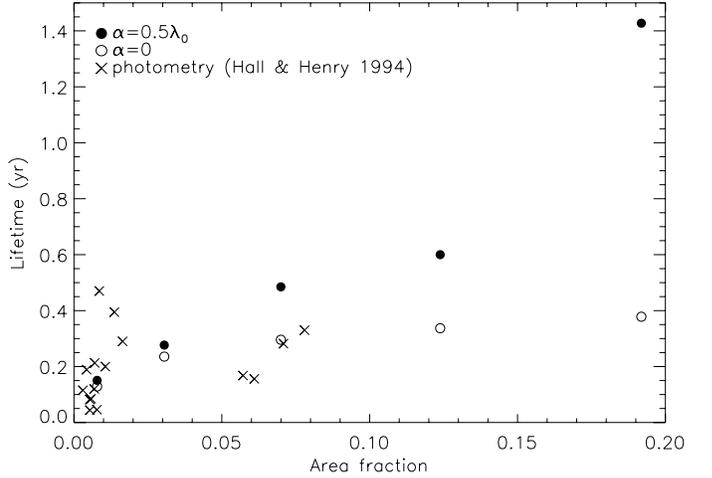}}
      \caption{Comparison of lifetime estimates of our simulations
      for different areas with photometric results of
      Hall~\&~Henry (\cite{hh94}).}
      \label{lifebigobs}
\end{figure}

Hall~\&~Henry (\cite{hh94}) give lifetimes for starspots of various
angular sizes based on their photometric light curve analysis. We compared
their lifetime estimates with the results of our simulations. We have
selected the main sequence G-K type single and binary stars (5 stars/systems
in total) from their sample. In Fig.~\ref{lifebigobs} we give the comparison.
Both sets of data are roughly consistent with each other.  
Hall \& Henry (\cite{hh94}) suggested two ``laws'' for starspot lifetimes,
based on long-term photometric observations on a sample of active main
sequence, subgiant, and giant stars of spectral types G, K, M. For very large spots, 
they assume that the lifetime is determined by shearing due to differential rotation, 
such that the spot lifetime is assumed to be 
inversely proportional to spot radius and differential rotation coefficient.
In the second relation, the lifetime is proportional to spot radius and stellar
radius; it is thought to apply for relatively small spots which have lifetimes
shorter than the disruption time scale of differential rotation. 
Since they do not give error
bars for spot sizes and lifetimes, we only take the data at the face value.
Their plots are logarithmic in both axes and the stellar sample covers stars
with very different radii and starspots probably at different latitudes (the
data does not include information on the latitudes of the spots). Different
stellar radii can cause a spread in lifetimes to within an order of
magnitude, considering the $R_{\star}^2$- dependence mentioned in
Sec.~\ref{sec:hr1099} and the weaker surface shear. Sources of error in the area
calculation based on brightness changes can also include the fact that there
is not only one big starspot but a cluster of spots with different sizes and 
lifetimes. Moreover, another spot might emerge
within the lifetime of the other, which would cause errors in both the area
and the lifetime. As a result, one cannot necessarily assume an individual 
coherent structure that is
having such a lifetime. According to the test models made by
Eker~(\cite{eker99}), a photometric signal with an amplitude of about 0.15
mag, for light curves which are accurate by $\pm 0.005$ mag in brightness
and by $\pm 0.005$ in the linear limb-darkening coefficient, the uncertainty of the
spot size is comparable to the size itself. 

We suggest that if the observed wave-like distortion in light curves 
can be caused by large spot conglomerates, differential rotation can 
disperse the
group of relatively small spots over a large area so that the initial
photometric signal (wave-like distortion seen in the light curves) would 
cease, but the individual spots can still survive without being detected,
having lifetimes proportional to sizes. Emergence of new spots 
may complicate the situation even further. Taking into account the fact
that unique solutions of light curves are in many cases not possible, the
rotational modulation of activity indicators as well as the patches in
Doppler maps can indicate different sub-structures for spot configurations
such as those considered in the simulations in Sect.~\ref{sec:starspots}.

Vogt et al.~(\cite{vogt99}), using Doppler images of \object{HR 1099} spanning 11
years, showed that the ``migrating photometric wave'' is not caused simply by
longitudinal migration of spots on a differentially rotating star but rather
to changes in the spatial distribution of a few spots that emerge and then
disappear, at the same mean longitude. The same authors also tracked two
long-lived spots and showed that they emerged at mid-latitudes and migrated upwards
to merge with the polar spot. They report a weak but
anti-solar differential rotation (poles rotating more rapidly than the
equator), contrary to our model assuming a solar-like differential rotation. 
However, one can obtain a long lifetime provided that the magnetic flux 
concentrates in a circular polar cap of radial field, regardless of 
the strength and the pattern of differential rotation. Our simulations 
also indicate that additional flux from tilted BMRs emerging at mid-latitudes 
can always be transported by the meridional flow to feed the polar cap flux. 
In Sec.~\ref{sec:bmrs} we have shown that a large tilted BMR emerging 
at an intermediate latitude is transported by meridional
flows and a solar-like differential rotation to form a polar spot. We have 
also shown that the follower-polarity flux from tilted BMRs emerging randomly in 
time and longitude, at
mid- to high latitudes can feed an already existing polar spot with additional flux. 

The simulations of surface magnetic flux transport presented in this paper 
let us draw the following conclusions.
\begin{itemize}

\item
The lifetime of a single, isolated bipolar magnetic region on a rapidly 
rotating star of solar radius is of the order of a few months, 
depending on the emergence latitude and the strength of surface shear. 

\item
The combined effects of diffusion and high emergence frequency are 
responsible for the short lifetimes of 
starspots and stellar magnetic regions at low and mid-latitudes. 

\item
Polar spots originate from the follower polarities of single or multiple 
tilted bipolar regions, emerging at 
mid- or high latitudes. A polar spot can be maintained by 
high-latitude emergence of new bipoles having the same polarity orientation. 

\item
Polar caps of evolved stars have lifetimes proportional to $R_{\star}^2$. 
For low-latitude regions, shorter lifetimes result from an increase of the 
diffusion rate due to differential rotation. 

\item
We find no clear indication as to whether large starspots are monolithic or 
clusters of smaller spots. The flux decay strongly depends on the 
details of the interaction between surface magnetic fields and 
large-scale flows, and the nature of turbulent diffusion. 

\end{itemize}

\begin{acknowledgements}
We thank Ingo Baumann for much help with the surface flux transport code.
\end{acknowledgements}

\end{document}